\documentclass
{aastex631}

\shorttitle{Internal Lorentz-Force Heating}
\shortauthors{Chyba and Hand}

\graphicspath{{./}{figures/}}

\begin{document}

\title{Internal-Current Lorentz-Force Heating of Astrophysical Objects}

\correspondingauthor{Christopher Chyba}
\email{cchyba@princeton.edu}

\author[0000-0002-6757-4522]{Christopher F. Chyba}
\affiliation{Department of Astrophysical Sciences  \\
Princeton University \\
4 Ivy Lane\\
Princeton NJ, 08544, USA}
\affiliation{Princeton School of Public and International Affairs  \\
Princeton University \\
221 Nassau Street\\
Princeton NJ, 08544, USA}

\author[0000-0002-3225-9426]{Kevin P. Hand}
\affiliation{Jet Propulsion Laboratory \\
California Institute of Technology \\
Pasadena CA, 91109 USA}

\begin{abstract}

Two forms of ohmic heating of astrophysical secondaries have received particular attention:  
unipolar-generator heating with currents running between the primary and secondary; 
and magnetic induction heating due to the primary's time-varying field.  Neither  
appears to cause significant dissipation in the contemporary solar system.
But these discussions have
overlooked heating derived from the 
spatial variation of 
the primary's field across the interior of
the secondary. 
This leads to Lorentz force-driven
currents around paths entirely internal to the secondary, with resulting ohmic heating.  We examine three ways to drive such currents, 
by the cross product of: (1) the secondary's azimuthal orbital velocity with the non-axially symmetric field of the primary; 
(2) the radial velocity (due to non-zero eccentricity) of the secondary with the primary's field; or (3) the out-of-plane velocity 
(due to non-zero inclination) with the primary's field. The first of these operates even for a spin-locked secondary whose orbit has zero eccentricity, 
in strong contrast to tidal dissipation. We show that Jupiter's moon Io today could dissipate about 600 GW (more than likely 
current radiogenic heating) in the outer hundred meters of its metallic core by this mechanism.
Had Io ever been at 3 jovian radii instead of its current 5.9, it could have been dissipating 15,000 GW.
Ohmic dissipation provides a  mechanism that could operate in any solar system to drive inward migration of secondaries that 
then necessarily comes  to a halt upon reaching a sufficiently close distance to the primary.

\end{abstract}


\section{Introduction} \label{sec:intro}

Two forms of ohmic (Joule) heating of astrophysical objects 
have been emphasized in the literature. The first, viewed in the reference frame  of
the rotating primary, is an
analog
to Lorentz-force-driven current flow and resulting ohmic dissipation in the Faraday disk
(\cite{Faraday1832, 
Munley2004, Chyba2015}).
Such unipolar  
heating has been explored as a dissipation mechanism 
for Jupiter's moons Io 
(\cite{Piddington1968, Goldreich1969, Drobyshevski1979, Colburn1980}) and Europa (\cite{Reynolds1983, Colburn1985}),
and Saturn's moon Enceladus (\cite{Hand2007}). It has also been
considered for
planetesimal heating
by the T-Tauri Sun (\cite{Sonnet1970}), and for astrophysical
binary systems (\cite{Laine2012}).
In the planetary-satellite instantiation of this hypothesis,
current flows in the ionosphere of the primary, down a flux tube
to the primary-facing equatorial region of the secondary, through the conducting secondary, and then back to the primary. 
But in the case of Jupiter's moon Io, the plasma likely shunts the circuit around Io itself, resulting in little
internal Joule heating (\cite{Colburn1980, Goertz1980, Russell2000, Saur2004}). At Europa, the current is 
limited by the resistance of the
ice shell overlying the conducting ocean; significant 
heating would require connecting the circuit to the ocean through cracks in the ice (\cite{Reynolds1983, Colburn1985}),
a possibility that should be reexamined now that possible plumes at Europa have apparently been observed
(\cite{Roth2014, Sparks2016}). At Enceladus, currents may be 
able to do just this, flowing through the
``tiger stripes" at the south pole, but even so the resulting Joule heating would provide $<1\%$ of the observed heat
flux (\cite{Hand2007}). 

A second form of ohmic heating featured in the literature
is magnetic induction heating due to eddy (Foucault)
currents 
driven by the primary's time-varying magnetic field.
Seen in the frame of the rotating (likely spin-locked) secondary, the primary's field varies with time due to the primary's rotation 
if there are off-axis components of its magnetic flux density ${\bf B}$,
or to variations in the field 
experienced by the secondary as it moves in an eccentric or inclined orbit. 
This model in effect treats the secondary as sitting in the interior field of a giant solenoid with spatially constant
but temporally oscillating ${\bf B}$. 
We have presented analytical induction heating
formulae for each of these cases,
and find that such heating appears
negligible for satellites in our solar system (\cite{Chyba2021}). 
In the case of highly conducting spheres (such as for Fe or Fe-S 
cores of satellites), total heating is limited because the oscillating
magnetic field penetrates only about 
one skin depth $\delta$ into the conductor. For Fe or Fe-S cores of Io or Europa, for example, $\delta\approx 100$
m, so nearly all of the core remains unheated. In the case of
a low-conductivity spherical shell (perhaps
a low-conductivity magma or liquid water ocean), the field can penetrate the conductor deeply, but then the inductive reactance becomes very
large. 
This finding of insignificant induction heating 
for objects in our solar system is consistent with earlier conclusions based on numerical treatments or waveguide models for specific objects
 (\cite{Colburn1980, Simonelli1983,
Khurana1998}). 
Exoplanets close to certain types of host stars
might experience significant induction
heating, however
(\cite{Laine2008, Kislyakova2017}).

 These discussions have
overlooked an additional ohmic
heating mechanism, one that derives from the
spatial variation of 
the primary's field ${\bf B}$ through or across the interior of 
the secondary.  
(There is something of an analogy to tidal heating, which results from the variation of the primary's gravitational field through  the secondary.) 
This leads to Lorentz-force-driven
currents around paths entirely internal to the secondary, with resulting dissipation. 
Here we show that this effect can generate significant heating for at least one moon in our current solar system, 
and perhaps greater heating in the past. It seems likely that analogous dissipation occurs in objects in extrasolar systems as well. 

\section{A New Mechanism} \label{sec:new}

The idea of this
proposed mechanism can be seen by considering the fundamental definition of electromotive force
(emf, or
$\varepsilon$), {\it viz.}
the work per unit charge done around a path $C$ due to the Lorentz force (e.g., \cite{Scanlon1969}):
$$\varepsilon=\oint_C ({\bf E} + {\bf v \times B}) \cdot {\bf dl}.\eqno(1)$$
Absent jump discontinuities  (e.g., \cite{Auchmann2014}) on the corresponding surface $S$  this becomes,
via Stokes' theorem and the Faraday-Maxwell  equation:
$$\varepsilon =\int_S[-\partial{\bf B}/\partial t+{\bf \nabla\times (v\times B)}] \cdot{\bf da}.\eqno(2)$$
We first work in a frame $K$ rotating with the primary. Consider a secondary orbiting in the equatorial plane of its primary 
in a circular orbit. Take the secondary to be synchronously rotating (spin-locked). Define coordinate systems with the usual conventions with 
origin at the center of the primary and $z$-axis along the primary's rotation axis. We then write the azimuthal velocity of the secondary as
$${\bf v_{\varphi}}=v_{\varphi}{\boldmath{\hat\varphi}}=\omega r\sin\theta{\boldmath{\hat\varphi}},\eqno(3)$$
using spherical coordinates $(r, \theta, \varphi)$ with
$\omega=\Omega-n$ the angular velocity of the secondary viewed from $K$, where $\Omega$ is the spin angular velocity of the 
primary and $n$ is the secondary's mean motion.
In $K$, $\partial{\bf B}/\partial t={\bf 0}$, and using ${\bf\nabla\cdot B=0}$
it is easy to show from Eq. (3) that (\cite{Chyba2016})
$${\bf \nabla\times (v_{\varphi}\times B)}=-\omega\left(\frac{\partial B_{r}}{\partial\varphi}\hat r+\frac{\partial B_{\theta}}{\partial \varphi}\hat\theta+\frac{\partial B_{\varphi}}{\partial \varphi}\hat \varphi\right)=-\omega\partial {\bf B}/\partial\varphi,\eqno(4)$$
which is {\bf 0}
for any axially symmetric ${\bf B}$.
Under these conditions, emf $= 0$ around $\it any$ interior path $C$ within the body of the secondary. So, for example, 
since Saturn's intrinsic magnetic field is azimuthally symmetric (\cite {Christensen2019}), the ${\bf v\times B}$ force cannot 
generate a non-zero emf around any interior path in a synchronously rotating satellite orbiting Saturn in a circular equatorial orbit.  
Similarly, the dipole, quadrupole, and octupole components of Jupiter's ${\bf B}$ field cannot generate an emf around any path 
in the interior of an analogous jovian satellite. But Eq. (4) also shows that an emf can be generated by those components of 
Jupiter's ${\bf B}$ field that vary azimuthally. Such components use the ${\bf v\times B}$ force to drive purely internal currents, 
even for ${\bf v}$ given by Eq. (3). The resulting energy dissipation (due to ohmic heating) operates even for spin-locked  secondaries with 
obliquity and orbital eccentricity equal to zero.  This contrasts with dissipative heating (and resulting orbital evolution) due to tidal 
effects: tidal dissipation within the secondary is zero for a spin-locked secondary in a circular orbit with zero obliquity (e.g., \cite{Chyba1989}).

But what about charge redistribution within the orbiting body?
We might expect the
 ${\bf v\times B}$ force to drive electron redistribution until the 
 resulting electrostatic field ${\bf E}$
perfectly cancels the ${\bf v\times B}$ field, so that ${\bf E=-v\times B}$ everywhere within the conductor, 
guaranteeing emf $=0$ in Eq. (1). This is true for many simple examples of conductors moving through magnetic fields 
(\cite{Lorrain1998}). 
The charge redistribution occurs extremely rapidly, on a classical  relaxation timescale 
$\tau_e\sim\epsilon_0/\sigma\approx 10^{-11}$ (1 ${\rm S~m}^{-1}/\sigma)$ s   (\cite{Redzic2004}), where $\sigma$ is 
electrical conductivity and $\epsilon_0$ vacuum permittivity.
In highly conducting metals the relaxation time is given by the electron collision timescale
$\tau_c\sim 10^7\tau_e$, or $\sim 10^{-11}$ s
 (\cite{Gutmann1974}).
In either case, charge would seem to redistribute 
rapidly and continuously 
to maintain ${\bf E = - v\times B}$, so that emf = 0 by Eq. (1) always. 
However, this argument fails when ${\bf \nabla\times(v\times B)\neq 0}$
(\cite{Chyba2016, Chyba2020}), because the electric field of a static charge distribution may always be written as a 
potential of a scalar field: ${\bf E = -\nabla}V$.
But since ${\bf \nabla\times \nabla} V{\bf =0}$ always, the equation
${\bf E} = -\nabla V = - {\bf v\times B}$ can hold only if
${\bf \nabla}\times ({\bf v\times B})=0$, which is violated in Eq. (4) for any ${\bf B}$ that varies with $\varphi$. 
Charge redistribution cannot stop a current from flowing in this case. If  ${\bf B}$ can be written as ${\bf B=B_0+B}(\varphi)$, 
where ${\bf B_0}$ is independent of $\varphi$ over $S$, electron redistribution will cancel the ${\bf v\times B_0}$ component. 
But this has no effect on the emf, because this component would integrate to ${\bf 0}$ around $C$ in Eq. (1) regardless.

Next we consider orbits for which $e\neq 0$.
A secondary orbiting with $e\neq 0$ has a component of its velocity radially toward or away from its primary, 
varying with the true anomaly $f$ around its orbit:
$${\bf v_r}=v_{r}{\boldmath{\hat r}}=n ae(1-e^2)^{-1/2}\sin f~{\boldmath{\hat r}},\eqno(5)$$
for semimajor axis $a$   (\cite{Murray1999}). Because this velocity varies with position around the secondary's elliptical orbit, 
and is independent of the primary's rotation, even in the frame $K$ the relevant angular velocity is $n$, not
$\omega$.
(One way to see this is to imagine the special case of a primary with an axisymmetric field and a secondary in an eccentric orbit. The relevant frequency for, say, the skin depth in the secondary is n, no matter how fast or slowly the primary is rotating.) 
 We find:
$${\bf \nabla\times (v_r\times B)}=\frac{v_r}{r}\left[\frac{1}{\sin\theta}\frac{\partial (B_{\theta}\sin\theta)}{\partial\theta}\hat r-\frac{\partial (rB_{\theta})}{\partial r}\hat\theta-\frac{\partial (rB_{\varphi})}{\partial r}\hat \varphi\right],\eqno(6)$$
so ${\bf v_r}$ can lead to emf generation even for axially symmetric primary ${\bf B}$ fields. 

 Finally we consider orbits for which inclination $i\neq 0$, and show that  these orbits, too, can generate electrical heating 
 via the axisymmetric dipole field (as well as, of course, via other components of the field).  A satellite orbiting with $i\neq 0$ has a 
 component of its velocity in the $\pm~\hat\theta$ direction, varying with $f$ around its orbit. We approximate this velocity by noting 
 that at apoapse, the secondary is at a height $z=a(1+e)\sin i$ above the primary's equatorial plane, whereas at periapse it is at a 
 height $z=-a(1-e)\sin i$ below the plane. Therefore in one-half an orbital period the secondary moves a vertical distance of $2a\sin i$, 
 giving it an average velocity in the $\hat\theta$ direction of
$${\bf v_{\theta}}=v_{\theta}{\boldmath{\hat \theta}}=(2/\pi)n a \sin i~{\boldmath{\hat \theta}}.\eqno(7)$$
Once again, the relevant angular velocity is $n$, not
$\omega$. We find:
$${\bf \nabla\times (v_{\theta}\times B)}=\frac{v_{\theta}}{r}\left\{-\frac{1}{\sin\theta}\frac{\partial (B_r\sin\theta)}{\partial\theta}\hat r+\left[\frac{1}{\sin\theta}\frac{\partial B_{\varphi}}{\partial \varphi}+\frac{\partial(rB_r)}{\partial r}\right]\hat\theta-\frac{\partial B_{\varphi}}{\partial \theta}\hat \varphi\right\},\eqno(8)$$
and an emf can be generated even if the primary field is axisymmetric. 

 All three cases considered here use the ${\bf v\times B}$ part of the Lorentz force in $K$ to drive currents around conducting 
 paths entirely interior to the secondary.
 The power $P$ dissipated   in the secondary is then given by
$$P=\overline{\varepsilon^2}R/Z^2, \eqno(9)$$ 
where $\overline{\varepsilon^2}$ is the square of the emf in Eq. (1), 
averaged around one orbit,
$Z=R+i\omega L$ is the conductor's impedance for the appropriate angular velocity , with $$Z^2=R^2+(\omega  L)^2,\eqno(10)$$
 and $R$ and $L$ the conductor's resistance and inductance, respectively. (For the case of ${\bf v_r}$ or  ${\bf v_{\theta}}$, $\omega$ in these expressions would be replced by $n$.) Values for $R$, $L$, and $Z$ have been previously 
 determined  for conducting spheres and spherical shells (\cite{Chyba2021}), geometries that roughly correspond to current 
 paths in secondaries' metallic cores, or spherical shells of conducting magma or liquid water oceans.

\section{Ohmic Heating for Conducting Spheres} \label{sec:ohmic}

We calculate the emf for the three cases considered here, on the assumption that the relevant part of the secondary in which
current flows is a solid sphere (for example, a conducting Fe or Fe-FeS core of a planetary satellite). We consider spherical shells 
(for example, magma or liquid water oceans) in Section 4.
We employ the usual magnetic field model for the primary field ${\bf B}$,
in terms of 
Schmidt quasi-normalized
associated Legendre polynomials with coefficients $g_l^m$ and $h_l^m$
of degree $l$ and order $m$ (e.g. \cite{Parkinson1983, Merrill1998}). Depending on the application, we make use of the
primary field's axisymmetric 
components (${\bf B^{\rm m=0}}$)  (Appendix A), or its nonaxisymmetric components (${\bf B^{\rm m\neq 0}}$) 
through second order (Appendix B). 

\subsection{Azimuthal velocities } 

We calculate the emf by Eq. (2) in $K$ for the case of azimuthal velocities ${\bf v_{\varphi}}$. In the frame $K$ rotating 
with the primary, ${\bf \partial B}/\partial t={\bf 0}$ not only for the axisymmetric components but for the nonaxisymmetric components 
as well. Then we can obtain the emf by integrating 
${\bf v_{\varphi}\times B}$ around relevant current paths. 
The ${\bf B^{\rm m=0}}$ components give ${\bf 0}$ to all orders by Eq. (4). 
To calculate the contribution of the ${\bf B^{\rm m\neq 0}}$ components, we first examine more carefully the argument that 
${\bf \partial B}/\partial t={\bf 0}$ in $K$. This certainly holds in frame $K$ for any imaginary curve $C$ orbiting 
the primary in empty space. However, consider the effects on the primary's field ${\bf B}$ at 
particular points in space as a conductor carrying $C$ passes through those points
during the conductor's 
orbit about the primary. Define the frame $K'$ to be the frame that orbits and rotates with the secondary.
In $K'$, the secondary sees an oscillatory time dependence due to the nonaxisymmetric field, meaning that the nonaxisymmetric field must fall off 
exponentially into the conductor with an e-folding distance given by the skin depth
 $$\delta=(2/\sigma \omega\mu)^{1/2},\eqno(11)$$
where we take magnetic permeability $\mu=\mu_0$ with $\mu_0=4\pi\times 10^{-7}$ H m$^{-1}$ the permeability of free space.
In general  $\mu=\mu_r\mu_0$. Setting the relative
permeability $\mu_r=1$ is clearly the right choice for rock or ice, but is also likely the correct choice for
 a satellite's iron core,
 because $\mu_r=1$ if the temperature of the core is above the Curie temperature
 $T_c = 1043$ K for iron, with little pressure dependence (\cite{Campbell2003}).

This same $\delta$ in Eq. (11) must be present in $K$ as well. 
For a planetary satellite, the relevant conducting sphere (of radius $r_o$) will presumably be made of Fe or Fe-FeS, for 
which $\sigma\sim 10^6$ S/m (\cite{Li2007, Silber2018}), and we will have $\delta\ll r_o$. So even in $K$, ${\bf B}$ changes with 
time as the conducting sphere orbits, in effect shielding successive regions of space from the ${\bf B^{\rm m\neq 0}}$ components of the field. 
However, 
this 
${\bf \partial B}/\partial t\neq{\bf 0}$ 
effect just drives the emf  we have previously calculated from an induction heating model in $K'$ that  
treats ${\bf B}^{\rm m\neq 0}$ as spatially constant but with a sinusoidal time dependence given by  $\varphi=\omega t$ (\cite{Chyba2021}). 
The emf in $K$ equals the emf $'$ in $K'$ to within a factor $(v/c)^2\sim 10^{-12}$ (\cite{Scanlon1969}). For $\delta\ll r_o$, the induction  
model gives $\varepsilon=\omega B^{1,1} 2\pi r_o\delta$ to leading order, smaller than the emf values we find in Eqs. (14) to (16) below by a 
factor $\sim\delta/r_o\sim 10^{-4}$, so its contribution to Eq. (2) can be ignored.

Current paths $C$ for our mechanism lie in the outermost-skin-depth layer of the conducting sphere
of the secondary. 
Because of the skin-depth effect, ${\bf B^{\rm m\neq 0}}$ at radius $s$ in the conducting sphere of radius $r_o$ falls 
off like ${{\bf B}^{\rm m\neq 0}(s)= {\bf B}^{\rm m\neq 0}(r_o)\exp{[-(r_o-s)/\delta}]}$ within the sphere. 
We employ a common approximation, 
treating ${\bf B}^{\rm m\neq 0}$ to penetrate with no attenuation an outer layer of thickness $\delta$ of the conducting 
sphere and to be {\bf 0} further within (\cite{Wouch1978}, \cite{Chyba2021}). 
An emf is generated around any path $C$ (with line element ${\bf dl}$) in this outermost layers of the conducting 
secondary for which $\oint_C{\bf  (v_{\varphi}\times B)\cdot dl}\neq 0$. By Eq. (3), we have
$${\bf v_{\varphi}\times B}=-v_{\varphi}B_{\theta}\hat r+v_{\varphi}B_r\hat\theta,\eqno(12)$$
so there are three orientations of current paths $C$  around which an emf may be driven: (a) paths 
in planes of constant $r$, 
extending from the nearside of the secondary to its farside, driven by the $\hat\theta$ term in Eq. (12); (b) 
paths in planes of constant $\theta$, driven by the $\hat r$ term in Eq. (12); and (c) paths in planes of constant $\varphi$, 
driven by both terms of Eq. (12). Ohmic heating results from currents running in each of these orthogonal sets of planes. 
Examples of these three paths are shown in Fig. 1.

Using Eqs. (B1) to (B3), we calculate the emf for path orientations (a), (b), and (c) by Eq. (1), 
integrating around a curve $\bar C$ that is the circumference of the average azimuthal current path of the relevant 
orientation in the conducting sphere (of radius $r_o$)
of the secondary, {\it viz.} $\bar C=\pi^2r_o/2$ (\cite{Chyba2021}),
corresponding to an annular radius 
$$\rho_o=\pi r_o/4.\eqno(13)$$
To make Eq. (1) analytically tractable, we choose paths of integration consisting of four legs, each 
leg locally parallel to $\hat r$, $\hat\theta$, or $\hat\varphi$; Fig. 1a illustrates this path for case (a), with the radial line 
from the primary to the secondary lying in the $\theta=\pi/2$ plane and along some arbitrary value $\varphi=\varphi_o$, which will 
subsequently be averaged over $2\pi$.
Then by Eqs. (1), (3), (12), and (B1) to (B3), and with ${\bf dl}=\hat r dr +\hat\theta rd\theta+\hat\varphi r\sin\theta d\varphi$,
$$\varepsilon=\oint_{\bar C_a}v_{\varphi}B_rrd\theta\approx \frac{\pi^2}{2}\omega r_o^2\left(\frac{R_p}{a}\right)^3\left[-g_1^1\sin\varphi_o+h_1^1\cos\varphi_o+\frac{3\sqrt{3}}{2}\left(\frac{R_p}{a}\right)(-g_2^2\sin 2\varphi_o+h_2^2\cos 2\varphi_o)\right],\eqno(14)$$
where  $v_{\varphi}B_2^1$ integrated to zero.
Similarly (see Fig. 1b), for cases (b) and (c) we find:
$$\varepsilon=-\oint_{\bar C_b}v_{\varphi}B_{\theta}dr\approx-\frac{\pi^2\sqrt{3}}{4}\omega r_o^2\left(\frac{R_p}{a}\right)^4(g_2^1\sin\varphi_o-h_2^1\cos\varphi_o);\eqno(15)$$
and (see Fig. (1c))
$$\varepsilon=\oint_{\bar C_c}v_{\varphi}(-B_{\theta}dr+B_rrd\theta)\approx\frac{\pi^2}{4}\omega r_o^2\left(\frac{R_p}{a}\right)^3\left[g_1^1\cos\varphi_o+h_1^1\sin\varphi_o+2\sqrt{3}\left(\frac{R_p}{a}\right)(g_2^2\cos 2\varphi_o+h_2^2\sin 2\varphi_o)\right].\eqno(16)$$
We sketch these calculations in Appendix C.

\begin{figure}
\epsscale{0.6}
\plotone{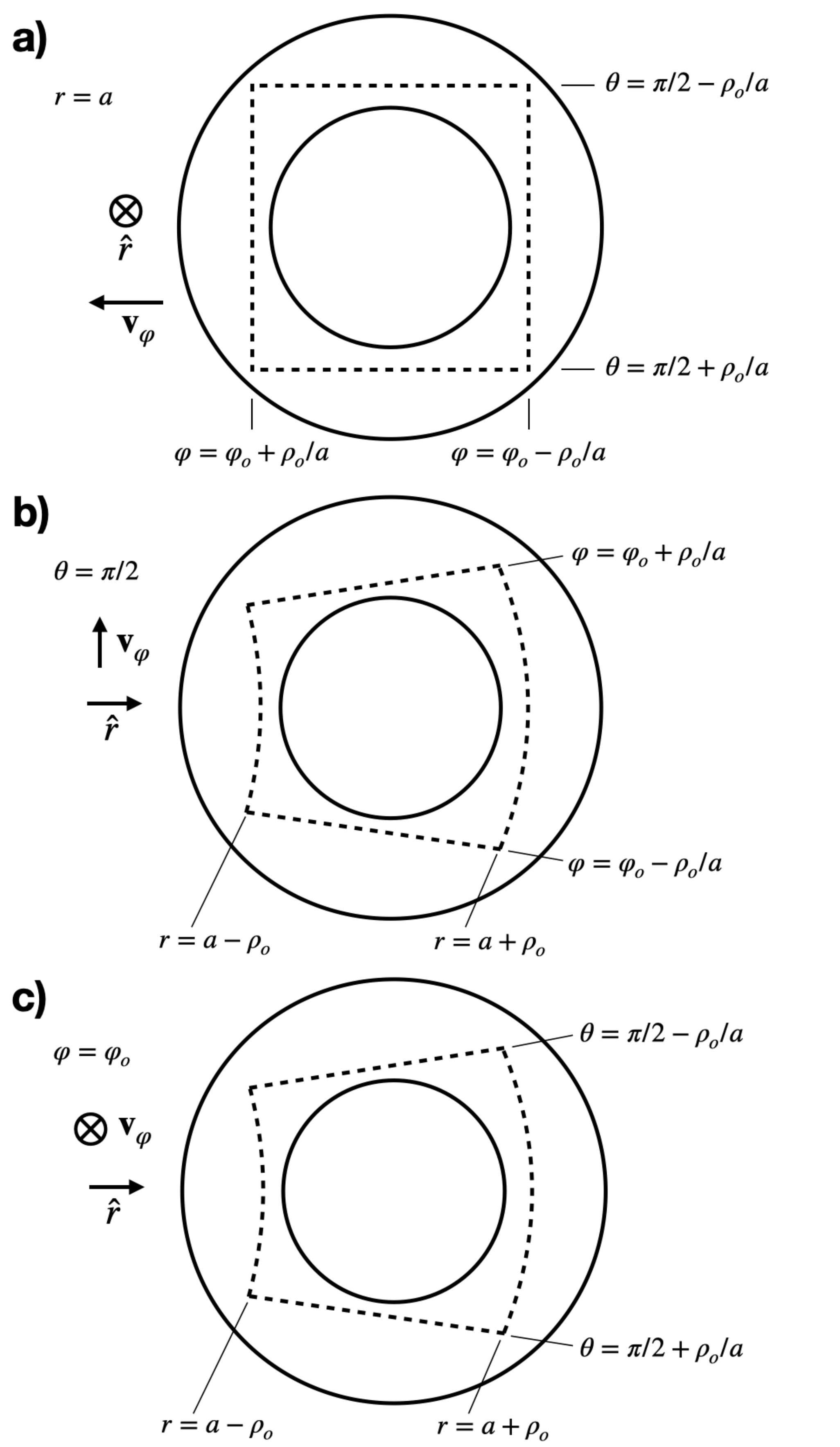}
\caption{Integration paths (dotted lines) used to calculate the line integrals in Eqs. (14), (15), and (16). The paths lie in the outermost skin depth of the conducting sphere, represented here by showing the paths to lie between concentric circles 
of radii $\rho_o-\delta$ and $\rho_o$. The radius $\rho_o$, defined in Eq. (13), is the radius corresponding to the average 
circumference for a current path for that orientation. In (a) segments are in the $\hat\theta$ and $\hat\varphi$ directions, 
in (b) the $\hat r$ and $\hat\varphi$ directions, and in (c) the $\hat r$ and $\hat\theta$ directions. Illustrations not to scale.} 
\label{fig:fig1}
\end{figure}

We now use Eqs. (9) and (10) to derive an expression for the power dissipated due to the emfs in Eqs. (14) to (16). 
The ohmic heating due to each path $C_a$, $C_b$, and $C_c$ is separately calculated and the three then summed for the total dissipation. 
 For a conducting sphere with $r_o\gg\delta$, we have (\cite{Chyba2021})
$$R_{\rm sphere}=\frac{\pi}{2\sigma\delta}~~~{\rm and}~~~
L_{\rm sphere}=
\frac{\pi\sqrt{3}}{4}\mu_0\delta,\eqno(17)$$
so that by Eq. (10),
$Z_{\rm sphere}^2=4R_{\rm sphere}^2$
and
$$\frac{R}{Z^2}=\frac{1}{\pi}\left(\frac{\sigma}{2\mu_0\omega}\right)^{1/2}.\eqno(18)$$
Therefore from Eqs. (9), (14) to (16), and (18), the power dissipated in the sphere due to its azimuthal velocity is, through second order:
$$P_{\rm sphere}(v_{\varphi})=\frac{\pi^3\sqrt{2}}{64}\left(\frac{\sigma}{\mu_0}\right)^{1/2}\omega^{3/2}r_o^4\left(\frac{R_p}{a}\right)^6 f(g_l^m,h_l^m),\eqno(19)$$
where 
$$f(g_l^m,h_l^m)= 5[(g_1^1)^2+(h_1^1)^2]+(R_p/a)^2\left\{3[(g_2^1)^2+(h_2^1)^2]+39[(g_2^2)^2+(h_2^2)^2] \right\}\eqno(20)$$
and we have averaged the products of the trigonometric functions of $\varphi_o$ over $2\pi$. 
The dissipation in Eq. (19), and resulting orbital evolution, is independent of the secondary's eccentricity. 
It is larger by a factor $\sim (r_o/\delta)^2$ than the corresponding induction heating (\cite{Chyba2021}). For a satellite with
an Fe or Fe-FeS core of radius $r_o\sim 10^3$ km, $(r_o/\delta)^2\sim 10^8$. 

We now examine ohmic heating that results from eccentric and inclined orbits.

\subsection{Radial velocities for eccentric orbits} 
Eq. (6) allows the calculation of ohmic heating due to eccentric orbits for the general case; 
here we restrict our attention to the special case $\theta=\pi/2$ that is likely to be relevant in many applications. 
We include only the dipole term (Appendix A), which for many primaries will be the leading component of the magnetic field. 
Including more terms is straightforward. We have
$${\bf v_r\times B}=
-v_rB_{\varphi}\hat\theta+v_rB_{\theta}\hat\varphi,\eqno(21)$$
and with Eq. (5) we calculate for the paths of cases (a) to (c) of Sec. 3.1:
$$\varepsilon(e\neq 0)=\oint_{\bar C_a}rv_{r}(-B_{\varphi}d\theta+B_{\theta}\sin\theta d\varphi)\approx
-\frac{\pi^2}{2}ne\sqrt{1-e^2}r_o^2\left(\frac{R_p}{a}\right)^3g_1^0\sin f,\eqno(22)$$
with an identical contribution from integrating around $C_b$, and zero contribution from the integral around $C_c$. 
Since $\sin^2 f$ averages to $1/2$ around the orbit, and with $r_o\gg\delta$ once again, by Eqs. (9) and (18), we have:
$$P_{\rm sphere}(e\neq 0)=\frac{\pi^3\sqrt{2}}{8}\left(\frac{\sigma}{\mu_0}\right)^{1/2}n^{3/2}e^2(1-e^2)r_o^4\left(\frac{R_p}{a}\right)^6\left(g_1^0\right)^2.\eqno(23)$$
Note that even in $K$, the relevant angular velocity here, including in the definition of $\delta$, is $n$, not $\omega$, since $v_r$ depends on $n$ and is independent of $\Omega$. (For intuition,  imagine a primary with an
axisymmetric field, in which
case it is clear that the
frequency relevant to
the skin
depth in the secondary is 
independent of the rotation
of the primary.)

\subsection{Velocities out of the plane for inclined orbits} 

Finally, we use Eq. (7) to calculate dissipation resulting from the secondary's out-of-plane velocity in an inclined orbit. 
Neptune's moon Triton, with its inclination of $157.3^\circ$ (\cite{NSSDC2014}) is our solar system's model for such a case. 
We display the result for the dipole term only; higher-order terms are readily calculated. We have
$${\bf v_{\theta}\times B}=
v_{\theta}B_{\varphi}\hat r-v_{\theta}B_{r}\hat \varphi,\eqno(24)$$
and with Eq. (5) we calculate for the paths of cases (a) to (c) in Sec. 3.1:
$$\varepsilon(i\neq 0)=\oint_{\bar C_a}v_{\theta}(B_{\varphi}dr-B_rr\sin\theta d\varphi)\approx
\pi n\sin i~ r_o^2\left(\frac{R_p}{a}\right)^3g_1^0,\eqno(25)$$
with an identical contribution from integrating around $C_b$, and zero contribution from $C_c$. With $r_o\gg\delta$ (again defined with
$n$, not $\omega$), we have:
$$P_{\rm sphere}(i\neq 0)=\pi\sqrt{2}\left(\frac{\sigma}{\mu_0}\right)^{1/2}n^{3/2}\sin^2i~r_o^4\left(\frac{R_p}{a}\right)^6\left(g_1^0\right)^2.\eqno(26)$$

\section{Ohmic Heating for Conducting Spherical Shells} \label{sec:shells}

We now present equations for ohmic heating for the three cases in Secs. 3.1, 3.2, and 3.3 above, except for conducting spherical 
shells (such as magma or liquid water oceans) rather than for solid spheres. We consider shells of outer radius $r_o$ and thickness $h$.

\subsection{Thick shells ($h\gg\delta$)} 

For a shell 
with $h\gg\delta$, the equations for $R$, $L$, and $Z$ are identical to those for the conducting sphere, so that Eq. (18) continues to hold
 (\cite{Chyba2021}). 
Therefore Eqs. (19), (22), and (25) are identical for conducting spheres and spherical shells. 

\subsection{Thin shells  ($h\ll\delta$)} 
 
In the opposite limit where $h\ll\delta$, we have (\cite{Chyba2021}$:$
$$R_{\rm shell}=\frac{\pi}{2\sigma h}~~~{\rm and}~~~
L_{\rm shell}=
\frac{\pi\sqrt{3}}{8}\mu_0 r_o,\eqno(27)$$
so that by Eq. (10):
$$\frac{R}{Z^2}=\frac{2\sigma h}{\pi}\left(1+\frac{3h^2r_o^2}{4\delta^4}\right)^{-1}.\eqno(28)$$
Therefore from Eqs. (9), (14) to (16), and (28), the power dissipated in the shell due to its azimuthal velocity is:
$$P_{\rm shell}^{h\ll\delta}(v_{\varphi})=\frac{\pi^3}{16}\sigma\omega^2r_o^4h\left(1+\frac{3h^2r_o^2}{4\delta^4}\right)^{-1}\left(\frac{R_p}{a}\right)^6 f(g_l^m,h_l^m),\eqno(29)$$
with $f(g_l^m,h_l^m)$ given by Eq. (20).
Power dissipation in the shell due to the radial velocity in an elliptical orbit is:
$$P_{\rm shell}^{h\ll\delta}(e\neq 0)=\frac{\pi^3}{4}\sigma n^2e^2(1-e^2)hr_o^4\left(1+\frac{3h^2r_o^2}{4\delta^4}\right)^{-1}\left(\frac{R_p}{a}\right)^6\left(g_1^0\right)^2;\eqno(30)$$
and dissipation due to the $\hat\theta$-velocity in an inclined orbit is:
$$P_{\rm shell}^{h\ll\delta}(i\neq 0)=2\pi\sigma n^2\sin^2i~r_o^4h\left(1+\frac{3h^2r_o^2}{4\delta^4}\right)^{-1}\left(\frac{R_p}{a}\right)^6\left(g_1^0\right)^2.\eqno(31)$$
Eqs. (29) to (31) are identical to within a small numerical factor to those found via a simple induction-heating model, 
which was shown  to generate negligible heating for satellites in our current solar system (\cite{Chyba2021}). In the limit 
where $\delta\gg h$, ${\bf B}^{\rm m\neq 0}$ is not attenuated by the conducting shell, so that in $K$ 
we have $\partial {\bf B}/\partial t={\bf 0}$ regardless of the secondary's orbital motion.

\section{Example: Ohmic Heating of Io} \label{sec:Io}

We illustrate the heating mechanism described here using two different interior models for Jupiter's moon Io. 
These will also serve as illustrative examples for ohmic heating of possible moons in extrasolar systems. 
A more comprehensive application of this model to other solar system satellites will be presented elsewhere. 
Io, a satellite of radius 1821.5 km, orbits a rotating 
Jupiter ($\Omega=1.76\times 10^{-4}$ s$^{-1}$) with $n=4.11\times 10^{-5}$ s$^{-1}$, $\omega=1.35\times 10^{-4}$ s$^{-1}$, $(R_p/a)=0.169$, $e=0.004$, and $i=0.04^{\circ}$ (\cite{NSSDC2014}). Jupiter's magnetic field has Schmidt coefficients $g_1^0=410244.7$ nT,
$g_1^1=-71498.3$ nT, $h_1^1=21330.5$ nT, $g_2^1=-56835.8$ nT, $h_2^1=-42027.3$ nT, 
$g_2^2=48689.5$ nT, and $h_2^2=19353.2$ nT (\cite{Connerney2018}).
We first use one possible interior model for Io (\cite{Schubert1986, Davies2007}) that takes it to have an Fe-FeS core of radius $r_o=950$ km with 
$\sigma=1\times 10^6$ S m$^{-1}$, appropriate to FeS at 
temperature 1900 K and pressure 6 GPa (\cite{Li2007}), approximately correct for 
the pressure and temperature at the upper boundary of Io's core. Liquid Fe at these pressures 
also has $\sigma\approx 1\times 10^6$ S m$^{-1}$ (\cite{Silber2018}).
(An alternate end-member Fe-core model 
for Io (\cite{Davies2007}) would have $r_o=650$ km.)
With these values, ${\bf B}$ has a skin depth
$\delta=(2/\sigma\omega\mu_0)^{1/2}=110 {\rm ~m}$ into the core.
Io's metallic core is overlain by a rock mantle of outer radius $1791$ km and thickness 
$h=840$ km (\cite {Davies2007}). It is unclear whether or not this mantle contains a liquid magma ocean (\cite{Khurana2011}, \cite{Bierson2016},  \cite{Blocker2018}).

The electrical conductivity of the mantle is unknown (e.g. \cite{Colburn1980}, \cite{Khurana2011}), 
and depends {\it inter alia} on the uncertain presence of the magma ocean.
First consider the case where the conductivity of Io's mantle 
is $<10^{-2}~{\rm S}$ m$^{-1}$, too low to shield Io's core from
Jupiter's time-varying {\bf B} field. (Even were Io to have a fully- or partially-shielded metallic core at present,
our mechanism could be of interest to an early Io prior to entering the Laplace resonance
(\cite{Yoder1979, Greenberg1982}), or to a 
variety of other moons or exo-moons.) In this model, Io's Fe-FeS core ($r_o=950$ km) 
is ohmically heated according to Eq. (19), which gives $P_{\rm Io}^{\rm core}(v_{\varphi})= 570$ GW. 
This is greater than the expected radiogenic heating for Io,
assuming chondritic composition (\cite{Cassen1982}).
The ohmic
 heating is concentrated 
in the outer 100 m of Io's Fe-FeS core,
with 
a power density $4.6\times 10^{-4}$ W m$^{-3}$, 
rather than being distributed throughout the lithosphere as for radiogenic heating. 
These results 
could affect heating profiles for interior models of Io (e.g. \cite{Bierson2016}),
and resulting physical conclusions. Six hundred gigawatts of ohmic heating is  $<1\%$ of Io's observed heat flow $1\times 10^{14}$ W, 
attributed to tidal dissipation
(\cite{Lainey2009,Veeder2012}).

Early Io would have rapidly become spin-locked (on a timescale $\sim 10^3$ yr) and its orbit circularized 
(on a timescale $\sim 10^7$ yr) (\cite{Murray1999}), after
which
 tidal dissipation in Io ceased until Io entered into resonance with Europa ({\cite{Yoder1979}), 
 unless this resonance were somehow primordial (\cite{Greenberg1982}). But even after spin-locking and orbit circularization, ohmic heating
 would have persisted and could have been high:
 If Io were closer to Jupiter in the past, ohmic heating would have increased like $ a^{-6}(\Omega-n)^{3/2}$. 
 For example, a spin-locked Io at $3R_p$ with $e=0$ would have experienced 15,000 GW of ohmic heating,
 dissipation that could  be important to understanding Io's thermal and orbital history.

By contrast, consider a second interior model in which a more conducting magma mantle shields the Fe-FeS core 
due to the skin effect, i.e. a mantle for which $h\gg\delta$. In this case, Eq. (19) again applies, with
$r_o$ the radius of the mantle.
The conductivity of the mantle in this scenario is uncertain but for illustration we take it to be at the upper end of 
plausible ultramafic rock melts, $\sigma=5$ S m$^{-1}$ (\cite{Khurana2011}). This value is about the same as that for a 
salty ocean (say on a Europa-like world). Then
$P_{\rm Io}^{\rm mantle}(v_{\varphi})=15$ GW, much smaller than radiogenic heating. 

\section{Orbital Evolution} \label{sec:orbits}

 Had Io experienced its current possible 570 GW of ohmic heating throughout the history of the solar system, 
 a total of $1\times 10^{29}$ J would have been dissipated in Io over that time, $\sim 1\%$ of Io's current orbital energy.
Nevertheless, ohmic heating might have been important to Io's orbital history, especially were Jupiter's tidal quality factor $Q_J$ large. 
As just noted, in an elegant scenario for the evolution of Io, Europa, and Ganymede into their three-body 
Laplace resonance (\cite{Yoder1979}), early Io rapidly despun, its orbit circularized, and its tidal dissipation ceased. 
Io then evolved outward in its orbit due to a jovian off-radial tidal bulge raised by Io. The resulting torque expanded  
Io's orbit faster than Europa's. Once Io entered into resonance, its eccentricity  increased, which in turn drove (and drives) 
tidal heating, possibly runaway melting, and perhaps
now even inward migration
out of the resonance
(\cite{Lainey2009}).
The characteristic timescale $\tau_p$ for orbital expansion due to torques from planet tides (tides
raised on Jupiter by Io)
is just (e.g. \cite{Chyba1989, Murray1999}):
$$\tau_p=\frac{2}{39}\left(\frac{M_p}{G}\right)^{1/2}\left(\frac{Q_p}{k_p}\right)\frac{a^{13/2}}{m_sR_p^{5}},\eqno(32)$$
where in our example $M_p$ is Jupiter's mass, $m_s$ the mass of Io, and we take $k_p=1/2$ to be the fluid Love number for 
Jupiter (\cite{Peale1979, Yoder1979}).

As with tidal dissipation in the secondary, ohmic dissipation comes out of the secondary's orbital energy $E_{\rm orb}$, 
so acts to decrease the semimajor axis of the orbit according to:
$$P_{\rm sphere}(v_{\varphi})=\dot E_{\rm orb}=\frac{GM_pm_s}{2a^2}\dot a,\eqno(33)$$}
giving a timescale for orbital contraction due to ohmic dissipation:
$$\tau_{\rm ohm}\equiv\frac{a}{\dot a}=\frac{GM_pm_s}{2aP_{\rm sphere}(v_\varphi)}.\eqno(34)$$
Using Eqs. (32), (34), and (19), we can compare the timescale $\tau_p$ for orbital expansion due to tides 
raised on the primary (in this case Jupiter), to the timescale $\tau_{\rm ohm}$ for orbital contraction due to ohmic 
dissipation in the secondary (in this case Io), and find that
$$\frac{\tau_p}{\tau_{\rm ohm}}=\frac{\pi^3\sqrt{2}}{96}\left(\frac{\omega a}{G}\right)^{3/2}\left(\frac{\sigma}{\mu_0}\right)^{1/2}\left(\frac{r_o^2}{m_s}\right)^2\left(\frac{Q_p}{k_p}\right)\left(\frac{R_p^2}{M_p}\right)^{1/2}f(g_l^m,h_l^m),\eqno(35)$$
with $f(g_l^m,h_l^m)$ from Eq. (20).
For larger $a$, orbital contraction due to ohmic dissipation in the secondary becomes increasingly 
important relative to orbital expansion due to tides on the primary.  This sets a limit to how far out a secondary with 
ohmic dissipation in its core can migrate. This 
contrasts with the analogous ratio $\tau_p/\tau_s$, where $\tau_s$ is 
the timescale for orbital contraction due to tidal dissipation in the satellite: the ratio $\tau_p/\tau_s$ is independent of $a$ (\cite{Chyba1989}).

For contemporary Io,  $\tau_p/\tau_{\rm ohm}\approx1\times 10^{-7}~ Q_p$.
\cite{Yoder1979} argues that  $2\times 10^5<Q_p<2\times10^6$, consistent with other values derived from tidal 
evolution arguments that require the Galilean satellites
not to have been pushed too far away from Jupiter over the age of the solar system, but to have been pushed enough to have entered into resonance (\cite{Goldreich1966, Greenberg1982}). 
Some interior models of Jupiter suggest values of $Q_p$ as large 
as $10^9$ or higher (\cite{Greenberg1982, Wu2005}); $Q_p ~{^>_{\sim}}~ 1\times 10^7$ in Eq. (35) would  imply a contemporary Io with a contracting orbit due to 
ohmic dissipation alone.
Such a world would migrate
inward until the
dependence in Eq.(35) on 
$[(\Omega-n)a]^{3/2}$ brought contraction into balance with orbital expansion driven by Jupiter tides and migration came to a halt.

However, \cite{Lainey2009} 
have used astrometric 
observations of the Galilean
moons 
to argue that Io is 
evolving inward due to 
tidal dissipation in Io, and
find
$Q_p/k_p=(1.102\pm 0.203)
\times 10^{-5}$ for Jupiter. For $k_p=0.5$, this gives
$Q_p=4.5\times 10^4$, in which case orbital
evolution due to ohmic 
dissipation never dominates
outward migration driven by
tides on Jupiter. 

Regardless of tidal dissipation in the primary, Eq. (19) shows that inward orbital migration due to ohmic 
dissipation must stop when $n=\Omega$. One can imagine a system (perhaps early solar system or extrasolar) 
in which $\tau_p/\tau_{ohm}>1$  in Eq. (35). Then the secondary would migrate 
inward until it reached $a=[G(M_p+m_s)/\Omega^2]^{1/3}$, corresponding to $n=\Omega$.
 For Jupiter the corresponding jovicentric distance of $2.2 R_p$ is comparable to the Roche limit, so that 
 secondaries might be (and could in the past have been) altogether lost.
But for primaries with somewhat smaller values of $\Omega$, the semimajor axis at which 
inward migration stops could lie well outside the Roche limit. However, if the secondary were heated due to ohmic 
dissipation in its core sufficient to form a thick and conductive ($h\gg\delta$) magma or liquid water ocean, the 
core would then become largely shielded from ${\bf B}^{\rm m\neq 0}$, causing heating to drop by as much as several orders of 
magnitude, leading the secondary to turn around in its migration (as $\tau_p/\tau_{\rm ohm}>1$ in Eq. (35) changes  
to $\tau_p/\tau_{\rm ohm}<1$) and expand its orbit due to the subsequently dominant effects of torques from  tides on the primary. 
Migration histories for secondaries, and implied limits for $Q_p$, are complicated by these potential histories. 
For secondaries such as Triton with substantial inclinations, Eq. (25) means that  $\tau_p/\tau_{\rm ohm}$ has a 
very different dependence on $a$ than that found in Eq. (35); we will explore this case elsewhere.

\begin{acknowledgments}
We thank P. J. Thomas and
an anonymous referee for 
their reviews, and G. Z. McDermott for reference assistance. K.P.H. 
acknowledges support from the Jet Propulsion Laboratory, California Institute of Technology, under contract with NASA. 
C.F.C. acknowledges support from Princeton University. 
\end{acknowledgments}

\appendix

\section{Axisymmetric  magnetic flux density through second order} 

We make such frequent use of the magnetic flux density (${\bf B}$) components of the primary  
through second order that we display them here in the appendix, rather than ask the reader to derive them 
from the magnetic potential $U$ (with ${\bf B=-\nabla}U$) whenever they are needed. We use the usual model 
(e.g. \cite{Parkinson1983, Merrill1998}) with $U$
written in terms of 
Schmidt-normalized
associated Legendre polynomials with coefficients $g_l^m$ and $h_l^m$
of degree $l$ and order $m$.
Units are those of magnetic flux
density.

The
axisymmetric dipole 
(the lowest-order axisymmetric field) 
then has the components
$$B^{1,0}_r=2(R_p/r)^3g_1^0\cos\theta\eqno(A1a)$$
$$B^{1,0}_{\theta}=(R_p/r)^3g_1^0\sin\theta\eqno(A1b)$$
$$B^{1,0}_{\varphi}=0,\eqno(A1c)$$
where $R_p$ is the appropriate reference radius for the primary,
and the superscript $``{1,0}"$ labels these as components of the dipole  field. 

The axisymmetric quadrupole has components:
$$B^{2,0}_r=\frac{3}{2}(R_p/r)^4g_2^0(3\cos^2\theta-1)\eqno(A2a)$$
$$B^{2,0}_{\theta}=3(R_p/r)^4g_2^0\cos\theta\sin\theta\eqno(A2b)$$
$$B^{2,0}_{\varphi}=0,\eqno(A2c)$$

Obviously
${\bf B^{1,0}}$ and ${\bf B^{2,0}}$ have no $\varphi$ dependence.

\section{Non-axisymmetric  magnetic flux density through second order} 

By Eq. (4) all $g_l^0$ terms contribute {\bf 0} to Eq. (2). The first-order terms of degree one,
$g_1^1$ and $h_1^1$, are typically the leading off-axis terms, corresponding
to orthogonal dipoles lying in the equatorial plane. 
They are given by
$$B^{1,1}_r=2(R_p/r)^3(g_1^1\cos\varphi+h_1^1\sin\varphi)\sin\theta\eqno(B1a)$$
$$B^{1,1}_{\theta}=-(R_p/r)^3(g_1^1\cos\varphi+h_1^1\sin\varphi)\cos\theta\eqno(B1b)$$
$$B^{1,1}_{\varphi}=(R_p/r)^3(g_1^1\sin\varphi-h_1^1\cos\varphi).\eqno(B1c)$$

The components of order 2, degree 1 are:
$$B^{2,1}_r=3\sqrt{3}(R_p/r)^4(g_2^1\cos \varphi+h_2^1\sin \varphi)\cos\theta\sin\theta\eqno(B2a)$$
$$B^{2,1}_{\theta}=\sqrt{3}(R_p/r)^4(g_2^1\cos \varphi+h_2^1\sin \varphi)(\sin^2\theta-\cos^2\theta)\eqno(B2b)$$
$$B^{2,1}_{\varphi}=\sqrt{3}(R_p/r)^4(g_2^1\sin \varphi-h_2^1\cos \varphi)\cos\theta.\eqno(B2c)$$

Finally, the order 2, degree 2 components are:
$$B^{2,2}_r=\frac{3\sqrt{3}}{2}(R_p/r)^4(g_2^2\cos 2\varphi+h_2^2\sin 2\varphi)\sin^2\theta\eqno(B3a)$$
$$B^{2,2}_{\theta}=-\sqrt{3}(R_p/r)^4[(g_2^2\cos 2\varphi+h_2^2\sin 2\varphi)\cos\theta\sin\theta\eqno(B3b)$$
$$B^{2,2}_{\varphi}=\sqrt{3}(R_p/r)^4(g_2^2\sin 2\varphi-h_2^2\cos 2\varphi)\sin \theta.\eqno(B3c)$$

\section{Example emf calculation}  
Here we show explicitly how the path in Fig 1a allows the calculation of the Eq. (14) line integral. 
The segments in Fig 1a lie in the $\hat\theta$ and $\hat\varphi$ directions, so by Eq. (12), only the two 
segments in the $\hat\theta$ direction contribute to the integral. By Eq. (4), $B^{1,0}_r$ and $B^{2,0}_r$ contribute 
nothing. To a sufficient approximation, we take $r=a$ (the choice $r=a+\rho_o$, say,  introduces terms of higher order 
in $\rho_o/a$). Through second order, Eq. (14) then becomes:
$$\varepsilon=\omega a^2\left[\int_{\pi/2+\rho_o/a}^{\pi/2-\rho_o/a}d\theta~\sin\theta(B^{1,1}_r+B^{2,1}_r+B^{2,2}_r)_{\varphi=\varphi_0-\rho_o/a}+\int_{\pi/2-\rho_o/a}^{\pi/2+\rho_o/a}d\theta~\sin\theta(B^{1,1}_r+B^{2,1}_r+B^{2,2}_r)_{\varphi=\varphi_0+\rho_o/a}\right].\eqno(C1)$$
By Eq. (B2a), the $B^{2,1}_r$ terms integrate to zero. After integration, the use of sum and difference formulas and  
small-angle approximations for the trigonometric functions gives the result in Eq. (14). Eqs. (15) and (16) are calculated 
analogously, though all four segments contribute to the integral in Eq. (16). In Eq. (15), to a sufficient approximation we 
take $\theta=\pi/2$; in Eq. (16) we take $\varphi=\varphi_0$.

\bibliography{DCLorentzApJLett}{}

\begin{thebibliography}{}
\expandafter\ifx\csname natexlab\endcsname\relax\def\natexlab#1{#1}\fi
\providecommand{\url}[1]{\href{#1}{#1}}
\providecommand{\dodoi}[1]{doi:~\href{http://doi.org/#1}{\nolinkurl{#1}}}
\providecommand{\doeprint}[1]{\href{http://ascl.net/#1}{\nolinkurl{http://ascl.net/#1}}}
\providecommand{\doarXiv}[1]{\href{https://arxiv.org/abs/#1}{\nolinkurl{https://arxiv.org/abs/#1}}}

\bibitem[{Auchmann {et~al.}(2014)Auchmann, Kurz, \&
  Russenschuck}]{Auchmann2014}
Auchmann, B., Kurz, S., \& Russenschuck, S. 2014, IEEE Trans. Magnetics, 50,
  doi: 10.1109/TMAG.2013.2285402

\bibitem[{Bierson \& Nimmo(2016)}]{Bierson2016}
Bierson, C.~J., \& Nimmo, F. 2016, J. Geophys. Res. Planets, 121, 2211

\bibitem[{Bl\"ocker {et~al.}(2018)Bl\"ocker, Saur, Roth, \&
  Strobel}]{Blocker2018}
Bl\"ocker, A., Saur, J., Roth, L., \& Strobel, D.~F. 2018, J. Geophys. Res.
  Space Phys., 123, 9286

\bibitem[{Campbell(2003)}]{Campbell2003}
Campbell, W.~H. 2003, Introduction to Geomagnetic Fields (Cambridge Univ.
  Press)

\bibitem[{Cassen {et~al.}(1982)Cassen, Peale, \& Reynolds}]{Cassen1982}
Cassen, P.~M., Peale, S.~J., \& Reynolds, R.~T. 1982, in Satellites of Jupiter,
  ed. D.~Morrison (Univ. Arizona Press), 93--128

\bibitem[{Christensen {et~al.}(2019)Christensen, Dougherty, \&
  Khurana}]{Christensen2019}
Christensen, U.~R., Dougherty, M.~K., \& Khurana, K. 2019, Saturn in the 21st
  Century, ed. K.~H. Baines, F.~M. Flasar, N.~Krupp, \& T.~Stallard (Cambridge
  Univ. Press), 69--96

\bibitem[{Chyba \& Hand(2016)}]{Chyba2016}
Chyba, C.~F., \& Hand, K.~P. 2016, Phys. Rev. Applied, 6, 014017

\bibitem[{Chyba \& Hand(2020)}]{Chyba2020}
---. 2020, Phys. Rev. Applied, 13, 028002

\bibitem[{Chyba {et~al.}(2015)Chyba, Hand, \& Thomas}]{Chyba2015}
Chyba, C.~F., Hand, K.~P., \& Thomas, P.~J. 2015, Am. J. Phys., 83, 72

\bibitem[{Chyba {et~al.}(2021)Chyba, Hand, \& Thomas}]{Chyba2021}
---. 2021, Icarus, 360, doi.org/10.1016/j.icarus.2021.114360

\bibitem[{Chyba {et~al.}(1989)Chyba, Jankowski, \& Nicholson}]{Chyba1989}
Chyba, C.~F., Jankowski, D.~G., \& Nicholson, P.~D. 1989, Astron. Astrophys.,
  219, L23

\bibitem[{Colburn(1980)}]{Colburn1980}
Colburn, D.~S. 1980, J. Geophys. Res., 85, 7257

\bibitem[{Colburn \& Reynolds(1985)}]{Colburn1985}
Colburn, D.~S., \& Reynolds, R.~T. 1985, Icarus, 63, 39

\bibitem[{Connerney {et~al.}(2018)Connerney, Kotsiaros, Oliverson, Espley,
  Joergensen, Roergensen, Merayo, Herceg, Bloxham, Moore, Bolton, \&
  Levin}]{Connerney2018}
Connerney, J. E.~P., Kotsiaros, S., Oliverson, R.~J., {et~al.} 2018, Geophys.
  Res. Lett., 45, 2590

\bibitem[{Davies(2007)}]{Davies2007}
Davies, A.~G. 2007, Volcanism on Io (Cambridge Univ. Press)

\bibitem[{Drobyshevski(1979)}]{Drobyshevski1979}
Drobyshevski, E.~M. 1979, Nature, 282, 811

\bibitem[{Faraday(1832)}]{Faraday1832}
Faraday, M. 1832, Phil. Trans. R. Soc. Lond., 122, 125

\bibitem[{Goertz(1980)}]{Goertz1980}
Goertz, C. 1980, J. Geophys. Res., 85, 2949

\bibitem[{Goldreich \& Lynden-Bell(1969)}]{Goldreich1969}
Goldreich, P., \& Lynden-Bell, D. 1969, Astrophys. J., 156, 59

\bibitem[{Goldreich \& Soter(1966)}]{Goldreich1966}
Goldreich, P., \& Soter, S. 1966, Icarus, 5, 375

\bibitem[{Greenberg(1982)}]{Greenberg1982}
Greenberg, R. 1982, in Satellites of Jupiter, ed. D.~Morrison (Univ. Arizona
  Press), 65--92

\bibitem[{Gutmann \& Borrego(1974)}]{Gutmann1974}
Gutmann, R.~J., \& Borrego, J.~M. 1974, IEEE Trans. Ant. Prop., 22, 635

\bibitem[{Hand {et~al.}(2011)Hand, Khurana, \& Chyba}]{Hand2007}
Hand, K.~P., Khurana, K., \& Chyba, C.~F. 2011, J. Geophys. Res., 116, E04010,
  doi:10.1029/2010JE003776

\bibitem[{Khurana {et~al.}(2011)Khurana, Jia, Kivelson, Nimmo, Schubert, \&
  Russell}]{Khurana2011}
Khurana, K.~K., Jia, X., Kivelson, M.~G., {et~al.} 2011, Scienceexpress,
  10.1126/science.1201425

\bibitem[{Khurana {et~al.}(1998)Khurana, Kivelson, Stevenson, Schubert,
  Russell, Walker, \& Polanskey}]{Khurana1998}
Khurana, K.~K., Kivelson, M.~G., Stevenson, D.~J., {et~al.} 1998, Nature, 395,
  777

\bibitem[{Kislyakova {et~al.}(2017)Kislyakova, Noack, Johnstone, Zaitsev,
  Fossati, Lammer, Khodachenko, Odert, \& G\"udel}]{Kislyakova2017}
Kislyakova, K.~G., Noack, L., Johnstone, C.~P., {et~al.} 2017, Nature Asron.,
  1, 878

\bibitem[{Laine \& Lin(2012)}]{Laine2012}
Laine, R.~O., \& Lin, D. N.~C. 2012, Asrophys. J. Lett., 745, doi:10.1088/0004

\bibitem[{Laine {et~al.}(2008)Laine, Lin, \& Shawfeng}]{Laine2008}
Laine, R.~O., Lin, D. N.~C., \& Shawfeng, D. 2008, Astrophys. J., 685, 521

\bibitem[{Lainey {et~al.}(2009)Lainey, Arlot, \"Ozg\"ur, \&
  Van~Hoolst}]{Lainey2009}
Lainey, V., Arlot, J., \"Ozg\"ur, K., \& Van~Hoolst, T. 2009, Nature, 459, 957

\bibitem[{Li {et~al.}(2007)Li, Gao, Zhang, He, Hao, Huang, C-L., Li, Li, Liu,
  \& Zou}]{Li2007}
Li, M., Gao, C.-X., Zhang, D.-M., {et~al.} 2007, Chin. Phys. Lett., 24, 54

\bibitem[{Lorrain {et~al.}(1998)Lorrain, McTavish, \& Lorrain}]{Lorrain1998}
Lorrain, P., McTavish, J., \& Lorrain, F. 1998, Eur. J. Phys., 19, 451

\bibitem[{Merrill {et~al.}(1998)Merrill, McElhinny, \& McFadden}]{Merrill1998}
Merrill, R.~T., McElhinny, M.~W., \& McFadden, P.~L. 1998, The Magnetic Field
  of the Earth (Academic)

\bibitem[{Munley(2004)}]{Munley2004}
Munley, F. 2004, Am. J. Phys., 72, 1478

\bibitem[{Murray \& Dermott(1999)}]{Murray1999}
Murray, C.~D., \& Dermott, S.~F. 1999, Solar System Dynamics (Cambridge:
  Cambridge Univ. Press)

\bibitem[{{National Space Science Data Center}(2014)}]{NSSDC2014}
{National Space Science Data Center}. 2014, Jupiter Fact Sheet.
\newblock \url{http://nssdc.gsfc.nasa.gov/planetary/factsheet/jupiterfact.html}

\bibitem[{Parkinson(1983)}]{Parkinson1983}
Parkinson, W.~D. 1983, Introduction to Geomagnetism (Scottish Academic Press)

\bibitem[{Peale {et~al.}(1979)Peale, Cassen, \& Reynolds}]{Peale1979}
Peale, S.~J., Cassen, P., \& Reynolds, R.~T. 1979, Science, 203, 892

\bibitem[{Piddington \& Drake(1968)}]{Piddington1968}
Piddington, J.~H., \& Drake, J.~K. 1968, Nature, 217, 935

\bibitem[{Red{\v z}i\'c(2004)}]{Redzic2004}
Red{\v z}i\'c, D.~V. 2004, Eur. J. Phys., 25, 623

\bibitem[{Reynolds {et~al.}(1983)Reynolds, Squyres, Colburn, \&
  McKay}]{Reynolds1983}
Reynolds, R.~T., Squyres, S.~Q., Colburn, D.~S., \& McKay, C.~P. 1983, Icarus,
  56, 246

\bibitem[{Roth {et~al.}(2014)Roth, Saur, Retherford, Strobel, Feldman, McGrath,
  \& Nimmo}]{Roth2014}
Roth, L., Saur, J., Retherford, K.~D., {et~al.} 2014, Science, 343, 171

\bibitem[{Russell \& Huddleston(2000)}]{Russell2000}
Russell, C.~T., \& Huddleston, D.~E. 2000, Adv. Space Res., 26, 1665

\bibitem[{Saur {et~al.}(2004)Saur, Neubauer, Connerney, Zarka, \&
  Kivelson}]{Saur2004}
Saur, J., Neubauer, F.~M., Connerney, J. E.~P., Zarka, P., \& Kivelson, M.~G.
  2004, in Jupiter: The Planet, Satellites and Magnetosphere, ed. F.~Bagenal,
  T.~E. Dowling, \& W.~B. McKinnon (Cambridge Univ. Press), 561--592

\bibitem[{Scanlon {et~al.}(1969)Scanlon, Henriksen, \& Allen}]{Scanlon1969}
Scanlon, P.~J., Henriksen, R.~N., \& Allen, J.~R. 1969, Am. J. Phys., 37, 698

\bibitem[{Schubert {et~al.}(1986)Schubert, Spohn, \& Reynolds}]{Schubert1986}
Schubert, G., Spohn, T., \& Reynolds, R.~T. 1986, in Satellites, ed. J.~A.
  Burns \& M.~S. Matthews (Univ. Arizona Press), 224--292

\bibitem[{Silber {et~al.}(2018)Silber, Secco, Yong, \& Littleton}]{Silber2018}
Silber, R.~E., Secco, R.~A., Yong, W., \& Littleton, J. A.~H. 2018, Nature Sci.
  Rep., 8

\bibitem[{Simonelli(1983)}]{Simonelli1983}
Simonelli, D.~P. 1983, Icarus, 54, 524

\bibitem[{Sonnet {et~al.}(1970)Sonnet, Colburn, Schwartz, \& Keil}]{Sonnet1970}
Sonnet, C.~P., Colburn, D.~S., Schwartz, K., \& Keil, K. 1970, Astrophys. Space
  Sci., 7, 446

\bibitem[{Sparks {et~al.}(2016)Sparks, Hand, McGrath, Bergeron, Cracraft, \&
  Deustua}]{Sparks2016}
Sparks, W.~B., Hand, K.~P., McGrath, M.~A., {et~al.} 2016, Astrophys. J., 829,
  121 doi:10.3847/0004

\bibitem[{Veeder {et~al.}(2012)Veeder, Davies, Matson, Johnson, Williams, \&
  Radebaugh}]{Veeder2012}
Veeder, G.~J., Davies, A.~G., Matson, D.~L., {et~al.} 2012, Icarus, 219, 701

\bibitem[{Wouch \& Lord(1978)}]{Wouch1978}
Wouch, G., \& Lord, A.~E. 1978, Am J. Phys., 46, 464

\bibitem[{Wu(2005)}]{Wu2005}
Wu, Y. 2005, Astrophys. J., 635, 688

\bibitem[{Yoder(1979)}]{Yoder1979}
Yoder, C.~F. 1979, Nature, 279, 767

\end{thebibliography}
\bibliographystyle{aasjournal}

\end{document}